\def\cm{{\rm\thinspace cm}}

\def\erg{{\rm\thinspace erg}}

\def\kpc{{\rm\thinspace kpc}}

\def\Msun{\hbox{$\rm\thinspace M_{\odot}$}}

\def\s{{\rm\thinspace s}}
\def\yr{{\rm\thinspace yr}}

\def\ergpcmsqps{\hbox{$\erg\cm^{-2}\s^{-1}\,$}}

\def\ergps{\hbox{$\erg\s^{-1}\,$}}

\def\pcmsq{\hbox{$\cm^{-2}\,$}}

\documentclass[usegraphicx]{mn2e}

\usepackage{fixltx2e}
\usepackage{mathptmx}

\include{defn}
\begin{document}

\title[High-z jets]{Do high redshift quasars have powerful jets?} \author[A.C. Fabian et al] {\parbox[]{6.5in}{{A.C. Fabian$^1$,
      S.A. Walker$^1$, A. Celotti$^{2,3,4}$, G. Ghisellini$^3$,
      P. Mocz$^5$, K.M. Blundell$^6$ and R.G. McMahon$^1$
    }\\
    \footnotesize
    $^1$ Institute of Astronomy, Madingley Road, Cambridge CB3 0HA\\
    $^2$ SISSA, via Bonomea 265, 34135, I-Trieste, Italy\\
    $^3$ INAF – Osservatorio Astronomico di Brera, via E. Bianchi 46,
    I-23807, Merate, Italy \\
    $^4$ INFN – Sezione di Trieste, via Valerio 2, I-34127, Trieste, Italy\\
    $^5$ Harvard University, Cambridge 02138, USA \\
    $^6$ Oxford University Astrophysics, Keble Road, Oxford, OX1 3RH\\
  }}
\maketitle
  
\begin{abstract}
  Double-lobed radio galaxies a few 100s of kpc in extent, like Cygnus
  A, are common at redshifts of 1 to 2, arising from some 10 per cent
  of the most powerful Active Galactic Nuclei (AGN). At higher
  redshifts they are rare, with none larger than a few 10s of kpc
  known above redshift $z\sim4$. Recent studies of the redshift
  evolution of powerful-jetted objects indicate that they may
  constitute a larger fraction of the AGN population above redshift 2
  than appears from a simple consideration of detected GHz radio
  sources. The radio band is misleading as the dramatic $(1+z)^4$
  boost in the energy density of the Cosmic Microwave Background (CMB)
  causes inverse Compton scattering to dominate the energy losses of
  relativistic electrons in the extended lobes produced by jets,
  making them strong X-ray, rather than radio, sources. Here we
  investigate limits to X-ray lobes around two distant quasars,
  ULAS\,J112001.48+064124.3 at $z = 7.1$ and SDSS\,J1030+0524 at
  $z=6.3$, and find that powerful jets could be operating yet be
  currently undetectable.  Jets may be instrumental in the rapid
  build-up of billion $\Msun$ black hole at a rate that violates the
  Eddington limit.
\end{abstract}

\begin{keywords}
black hole physics: accretion discs, X-rays: binaries, galaxies 
\end{keywords}

\section{Introduction}

Extended X-ray emission produced by inverse Compton scattering of CMB
photons in distant giant radio sources has been considered several
times (Felten \& Rees 1969; Schwartz 2002a; Celotti \& Fabian 2004;
Ghisellini et al 2013).  Relativistic electrons with Lorentz factor
$\Gamma\sim10^3$ upscatter CMB photons into the soft X-ray band
observed around 1 keV. The boost factor in CMB energy density
compensates for surface brightness dimming with redshift. The X-ray
phase is also expected to even last longer than the radio phase
(Fabian et al 2009; Mocz et al 2011).  Many examples (e.g. Carilli et
al 2002; Scharf et al 2003; Overzier et al 2005; Blundell et a; 2006;
Erlund et al 2006; Laskar et al 2010; Smail et a; 2012; Smail \&
Blundell 2013) of double-lobed sources simultaneously luminous in both
radio and X-rays are known across a wide range in redshift up to
$z\sim4$. The highest redshift example of a $>100\kpc$ scale
double-lobed structure is 4C23.56 (Blundell \& Fabian 2011) at $z =
2.5$. When inverse Compton losses are considered, the dearth of large
radio galaxies at higher redshifts does not necessarily indicate a
lack of high redshift jetted sources (Mocz et al 2013; Ghisellini et
al 2014).

At $z>7$ the energy density of the CMB is over 4000 times greater than
at the present epoch and very likely far exceeds that of the magnetic
field in any lobes produced by a jet, so making synchrotron radio
emission weak and undetectable at GHz frequencies.  Hard X-ray
emission detected from relativistically-beamed jets by the Burst Alert
Telescope (BAT, onboard Swift) implies that the ratio of powerful
radio-loud AGN to radio-quiet AGN strongly increases with redshift
(Ghisellini et al 2013, 2014).  The number density obtained when
beaming corrections are accounted for nearly matches that of all
massive dark matter haloes capable of hosting billion-solar mass black
holes above redshift 4 (Ghisellini et al 2013). It is therefore
possible that X-ray lobed sources produced by powerful jets exist, and
could even be common, at high redshift. In other words, high redshift
quasars may have powerful jets.

To test whether powerful high redshift quasars generate powerful jets,
we searched for linearly extended, diametrically-opposed X-ray lobes
around archival X-ray images of two distant quasars, the most extreme
object, ULAS\,J1120+0641 at $z=7.1$, and SDSS\,J1030+0524 at
$z=6.3$. Both have relatively deep XMM exposures (Page et al 2014 and
Farrah et al 2004, respectively). Schwartz (2002b) has previously
searched for X-ray jets around 3 $z\sim6$ quasars using short ($\sim
10$\,ks) Chandra images without any conclusive result. 

\section{The X-ray images}
\subsection{ULAS\,J1120+0641}
Neither the Chandra image (which is short, 16\,ks) nor the XMM-Newton
pn + MOS image data ( Page et al 2014; total exposure of 300 ks which
on reanalysis gives 120 ks of useable, low background, exposure) shows
any significant linear structure around the quasar (Fig. 1). A linear
structure which is $\sim 1$ arcmin long (projected length of 318 kpc
at $z=7.1$, corresponding to a true length of $>385\kpc$ if inclined
at $<60$~deg to our line of sight) is seen about one arcmin to the
WSW, centred on a weak point source. The structure either side of the
weak source is at a statistically significant flux above the
background (4.3 sigma), but we note smaller blobs (possibly sources)
at the same surface brightness level and suspect that it may be due to
source confusion. It is difficult to deduce a precise significance for
any weak extended structure but note that in this and other XMM images
we see no similar linear features. Where lines do appear they
correlate with chip gaps. The structure does not align with a chip gap
or anything on the exposure map. The linear feature is present below
about 1.5 keV and is best seen between 0.3-1 keV.  We shall take this
structure as representative of the upper limit of what is
undetectable.  No object consistent with high redshift is seen at the
position of the weak point X-ray source in HST imaging of the field
(Simpson et al MNRAS submitted).

\begin{figure}
  \centering
  \includegraphics[width=0.99\columnwidth,angle=0]{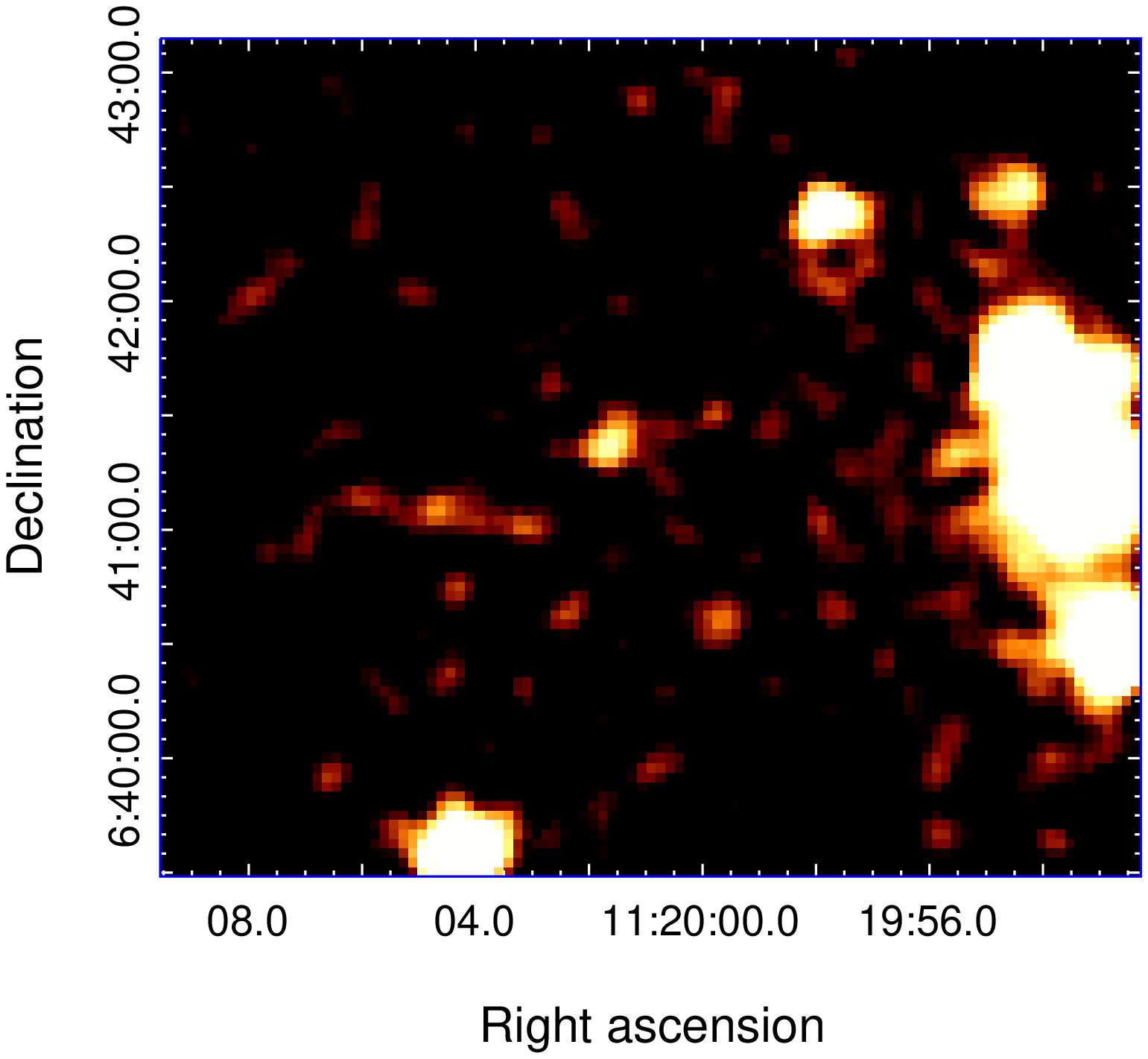}
  \includegraphics[width=0.99\columnwidth,angle=0]{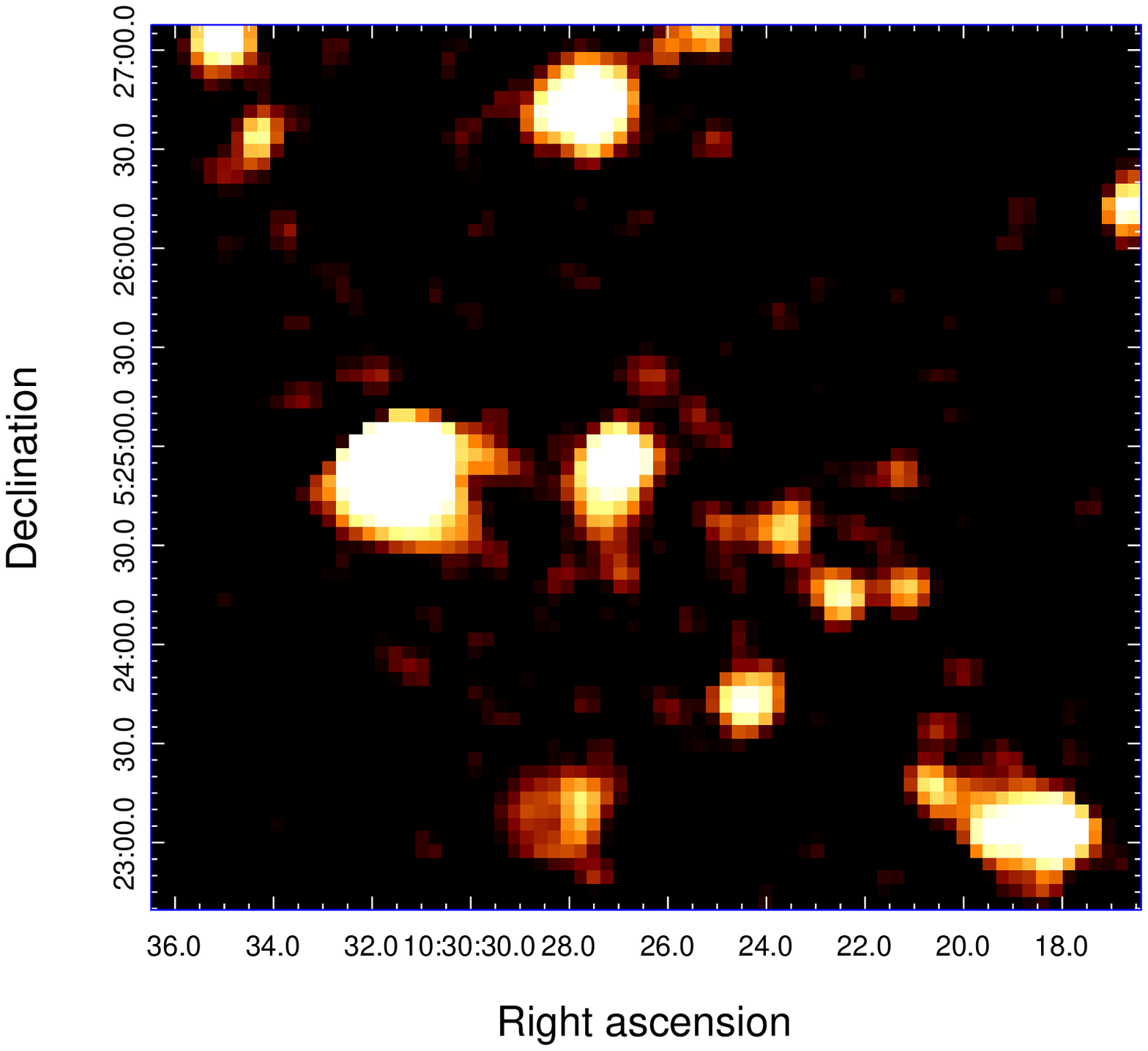}
  \caption{XMM images of ULAS\.J1120+0641 ($z=7.1$, top) and
    SDSS\,J1030+0524 ($z=6.3$, bottom). The energy band of both images
    is 0.3-1\,keV.}
\end{figure}

In Fig. 2 we show the regions used for the source and background
overlaid on the merged image and the merged exposure map. The merged
exposure map contains no features that could have caused the observed
lobe structure. For the 0.3-1 keV band, we find a total of 69 (48)
counts for the pn+MOS (pn) in the two regions (arms)on either side of
the weak source, shown as green boxes in Fig. 1.  The background level
estimated from the large region indicated by the dashed green box implies an
expectation of 41.5 (26.5) counts in the source boxes. The arms
therefore correspond to 27.5 (21.5) counts, or 4.3 (4.2) sigma, above
background, which is an observed flux of $2\times10^{-16} \ergpcmsqps$
in the 0.3-1 keV band. Correcting the flux for Galactic absorption
(hydrogen column density of $5\times10^{20} \pcmsq$, Kalberla et al
2005) gives an intrinsic 2-10 keV luminosity of
$2.2\times10^{44}\ergps$.  We henceforth use that value as our
detection limit on large linear structures, such as X-ray lobes.

No radio emission has yet been detected from ULAS\,J1120+0641 (Momjian
et al 2014) or its near vicinity (analogously to the $z \sim 2$ AGN
HDF130 that only has X-ray lobes, Fabian et al 2009; Mocz et al
2011). The lack of radio emission in this $z = 7.1$ quasar is
consistent with our modeling results if the magnetic field in the
lobes is about 10~$\mu$G., which is about a factor of 5 below
equipartition. Deeper lower frequency radio imaging could detect faint
radio emission from any lobes, if present.

\subsection{SDSS\,J1030+0524}
Farrah et al (2004) discuss SDSS\,J1030+0524 and its XMM data, which
has about 75~ks of useable data. The X-ray image is shown in Fig.~1
(lower) and shows some structures close to the quasar which are most
probably due to weak confused sources. We adopt the same upper limit
to that for ULAS\,J1120+0641, given the similarities in XMM exposure
time. The source was undetected in the radio band at 1.4~GHz by Petric
et al (2003). The arcmin size, double radio source one arcmin S of the
quasar shown in their 1.4GHz radio map is marginally coincident with
several weak X-ray sources. No further information or identification
has been made of that radio source.

\begin{figure}
  \centering
  \includegraphics[width=0.99\columnwidth,angle=0]{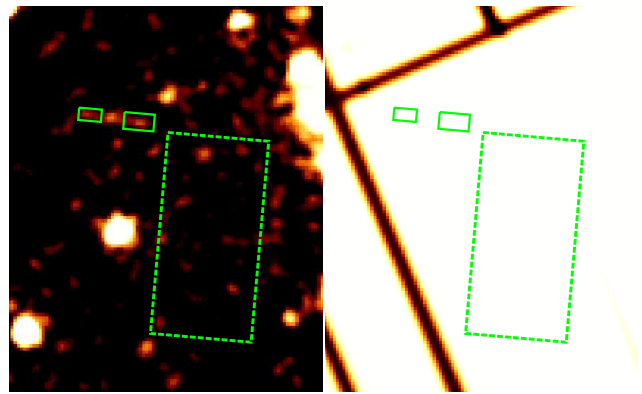}
  \caption{Fig.~2. Left: 0.3-1.0 keV merged MOS and PN image showing the
    spectral extraction regions for the lobes (green boxes) and the
    background (blue box). Right: The same region files overlaid on
    the merged exposure map, showing no structures in the exposure
    maps that could have erroneously caused the linear lobe feature
    around the quasar.}
\end{figure}

\section{Modelling}

The luminosity upper limit we have adopted is similar to that
predicted by recent models for jetted lobe evolution (Ghisellini et al
2014; Mocz et al 2013).  The properties of a jet powering a radio lobe
at $z=7.1$ are estimated using the analytic model developed in Mocz et
al (2013). The model treats the evolution of the hydrodynamics of the
radio lobe and derives the synchrotron emission (in the radio) and the
IC emission (in the X-rays) from it, also accounting for adiabatic
expansion energy losses. We fix the magnetic field to $10\mu$G and use
fiducial parameters for the injection spectrum and surrounding gas
profile (set [A] in Mocz et al (2011): the injection spectrum is given
by a power-law index 2.14 and Lorentz factors ranging between 1 to
$10^6$; the surrounding density profile has a powerlaw index of
1.5). We assume each lobe has length 185 kpc so that the jet-axis is
at 60 deg to our line of sight and model the observed radio lobe at
$z=7.085$ with a (very powerful) jet power of $Q=8\times10^{45}
\ergps$ per lobe.  The lobes each grow to 185 kpc in
$1.4\times10^7\yr$. The observed) X-ray power (1 keV) of a single lobe
given by the model at this point in the evolution is $1.1\times10^{44}
\ergps$ (Fig. 3). The observed radio power (1.4 GHz) is estimated to
be $1.8\times10^{41} \ergps$.

\begin{figure}
  \centering
  \includegraphics[width=0.99\columnwidth,angle=0]{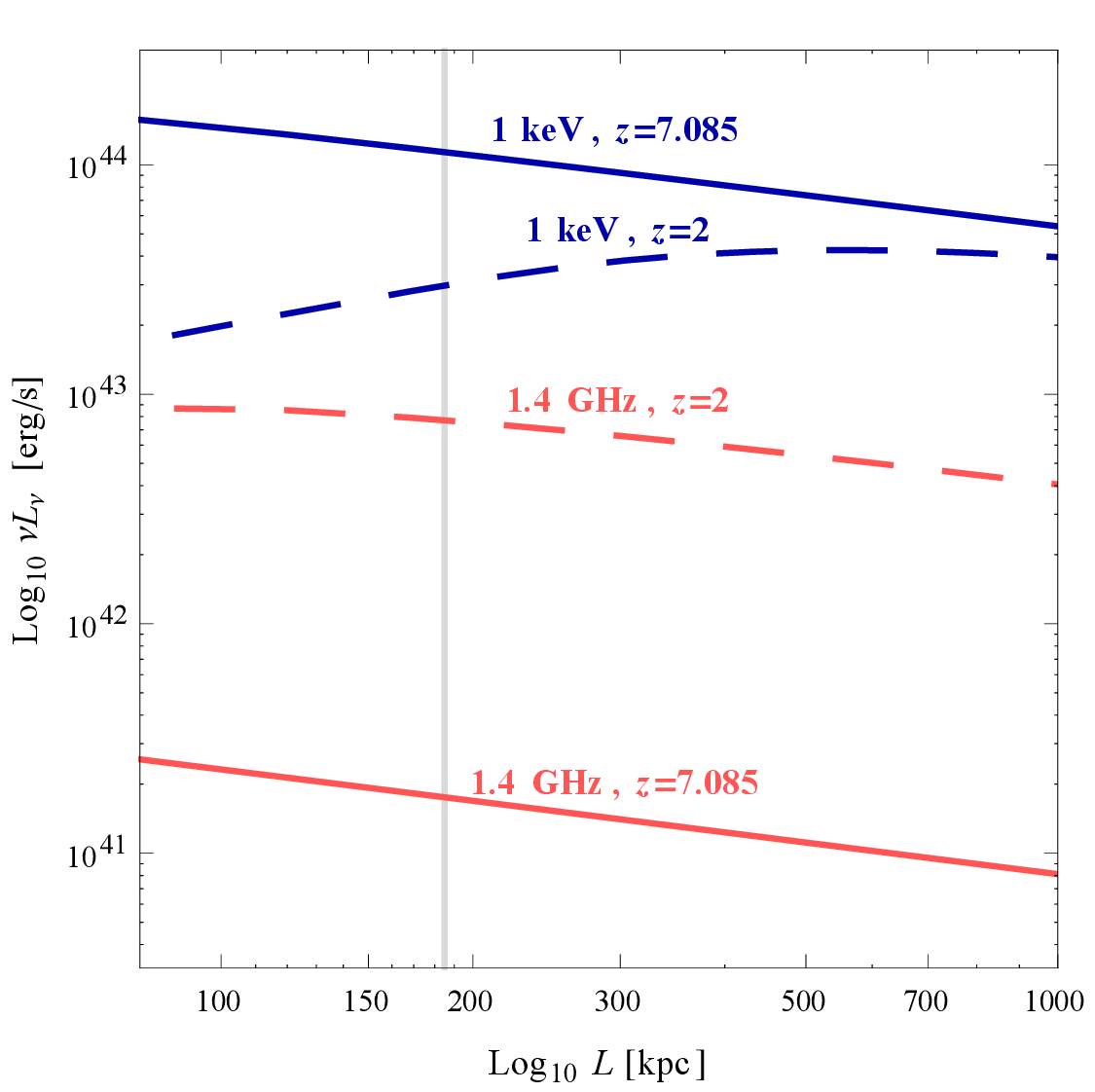}
  \caption{Fig. 3. Modelled observed X-ray and radio powers versus
    length of a single radio lobe at $z=7.085$ with jet power
    $Q\sim2\times10^{46} \ergps$ shown as solid lines. A lobe is also
    modelled at $z=2$ for comparison, and shown as dashed lines. At
    higher redshift, the energy density of the CMB is larger,
    resulting in a source that is brighter in its IC emission (X-ray)
    and lower in its synchrotron emission (radio) due to IC energy
    losses of the electrons. The length of each lobe is taken to be
    185 kpc which assumes that it is oriented at 60 deg to our line of
    sight. Leptonic jets are assumed; if protons are present the jets
    can carry 10 times more energy. }
\end{figure}

This thus requires a total jet power of $2\times10^{46}\ergps$ acting
for over $10^7\yr$ in order for each lobe to grow to longer than 185
kpc.  For an accretion efficiency of 0.1 the black hole will have
grown by $4\times10^7\Msun$ in this jet outburst. A longer outburst
leading to larger lobes. or several shorter outbursts would enable
this process to account for much of the black hole mass. If the black
hole is spinning rapidly then the accretion efficiency increases and
the accretion rate reduces. However this may be more than offset by a
significant proton content to the jets which can increase the jet
power tenfold and make the black hole grow much faster.

\section{Discussion}

We have shown that inverse Compton lobes produced by powerful jets
powered by the central quasar could be currently undectable despite
the fact that the jet power rivals that radiated by the quasar. This
becomes more extreme if the jet contains protons as well as
electrons. Jets could be the result of the extraction of rotational
energy of the black hole (Blandford \& Znajek 1977), but the accretion
disk must amplify the magnetic field needed for the extraction of the
black hole spin energy. This can correspond to the major energy
release of the accretion process which, it is widely assumed, powers
the quasar. If substantial energy is extracted from the accretion disc
magnetically, then the total power of the source could exceed the
Eddington limit, a constraint which applies only to the power in photons
(see discussion in Ghisellini et al 2013).  A billion $\Msun$ black
hole can thereby grow from stellar mass by accretion before redshift
7.1.

The upper limit to any X-ray lobes of ULAS\,J1120+0641 (and
SDSS\,J1030+0524) imply that the process of massive black hole growth
at large redshift could include powerful-jetted outflows. Much of the
radiation is due to inverse-Compton upscattering of CMB photons into
the X-ray band. The enormous mechanical energy of jetted lobes would
represent a formidable and fierce form of kinetic feedback on the
surrounding gas (Fabian 2012). This could explain why the galaxy hosts
of quasars at $z>3$ are compact (Szomoru et al 2013), and their group
and cluster gas have more energy than is explainable by gravitational
infall alone (Wu et al 2000; McCarthy et al 2012). Powerful jets are a
considerable source of energy, magnetic fields and cosmic rays to the
intergalactic medium\footnote{See Gopal-Krishna et al (2001) and Mocz
  et al (2011) for discussion on the impact of jetted lobes on the
  intergalactic medium at $1<z<4$.}and may therefore play a role in
reionization of the Universe at these early cosmic times. Our results
suggest that the high redshift Sky could be criss-crossed by X-ray
emitting jets and lobes from growing massive black holes. Deeper X-ray
imaging with Chandra and, in time, Athena can test this possibility.

\section*{Acknowledgements} 
We thank the referee for helpful comments.
This material is based upon work supported by the National Science
Foundation Graduate Research Fellowship (PM) under Grant
No. DGE-1144152. ACF and SAW thank the Science and Technology Research
Council. AC acknowledges the Institute of Astronomy for warm
hospitality.


\end{document}